A Unified View of Topological Phase Transition in Band Theory

Huaqing Huang[1,2], Feng Liu[1]

[1] Department of Materials Science and Engineering, University of Utah, Salt Lake City, Utah 84112, USA

[2] School of Physics, Peking University, Beijing 100871, China

Correspondence should be addressed to Feng Liu; fliu@eng.utah.edu

We develop a unified view of topological phase transitions (TPTs) in solids by revising the classical band theory with the inclusion of topology. Re-evaluating the band evolution from an "atomic crystal" [a normal insulator (NI)] to a solid crystal, such as a semiconductor, we demonstrate that there exists ubiquitously an intermediate phase of topological insulator (TI), whose critical transition point displays a linear scaling between electron hopping potential and average bond length, underlined by deformation-potential theory. The validity of the scaling relation is verified in various two-dimensional (2D) lattices regardless of lattice symmetry, periodicity, and form of electron hoppings, based on a generic tight-binding model. Significantly, this linear scaling is shown to set an upper bound for the degree of structural disorder to destroy the topological order in a crystalline solid, as exemplified by formation of vacancies and thermal disorder. Our work formulates a simple framework for understanding the physical nature of TPTs with significant implications in practical applications of topological materials.

1. **Introduction**
Band theory is one of the most important developments of condensed matter and material physics, which underlines the working principle of modern electronic and optoelectronic devices. It is well known that isolated atomic levels would spread to form energy bands when atoms were brought together to form a solid [1], which provides a general band evolution process to understand metal, semiconductor and insulator states. However, this classical *textbook* picture is incomplete and must be revised in light of recent emergence of band topology. In general, an insulating solid is a NI if it can be adiabatically connected (without gap closure or band inversion) to the "atomic limit"; otherwise, it is a TI [2]. Namely, the TPT from a NI to a TI requires a band inversion between conduction and valence bands [3]. This implies that ubiquitously, an intermediate TI phase should appear during the band evolution process when the spin-orbit coupling (SOC) effect is included. A *textbook-revised* generic band evolution diagram is shown schematically in Figure 1 (see also Figure S1 in Supplementary Material). Starting from the atomic limit (a NI), $s$ and $p$ levels are initially separated by a trivial charge gap $\Delta_{sp}$. The $p$ level splits due to SOC effect. By reducing the average bond length ($L$), the orbital levels spread to form individual bands with a finite bandwidth $W$. Consequently, the charge gap reduces and closes eventually to realize an $s$-$p$ band inversion. Then the SOC effect reopens an energy gap with nontrivial topology, leading to a NI-to-TI TPT. Further reducing the average bond length to overcome the SOC gap will drive the system into a gapless phase before reaching a semiconducting phase with strong $s$-$p$ hybridization. Therefore, a fundamental question is when and how the TPT occurs within the framework of band theory?

The study of TPT dates back to 1970s when phenomena in quantum states of matter, such as the quantum Hall effect [4] and superfluid phase transitions in 2D [5], were explained using the mathematical concepts of topology [6–8]. These pioneering works have since paved the way for the introduction of many new topological states such as quantum anomalous/spin Hall effects [9–11], 3D TIs [12, 13] and topological superconductors [14], and revolutionized electron band theory [15]. Generally, topological states are insensitive to a smooth change of material parameters unless the system passes through a TPT. So far, various TIs with different band inversion mechanisms have been theoretically proposed and/or experimentally verified, either periodic [16, 17] or aperiodic [18–22]. The critical condition of TPTs are determined on a case-by-case basis, but there is no unified view on the TPT among different systems. Fundamentally, it remains unclear how a TI is non-adiabatically connected to the "atomic limit" during the gap closure process?

It is important to recognized that conventional phases transitions, as described by Landau theory of spontaneous symmetry breaking, exhibit a universal scaling relation of criticality. Differently, TPTs, involving no symmetry breaking, are characterized by a sudden change of topological invariants with a continuously changing system parameter. Thus, one does not expect a form of universality to be associated with TPTs. Surprisingly, however, we discover a linear scaling relation between electron hopping potential and average bond length within the framework of band theory, which is commonly applicable to TPTs in different systems albeit with different slopes, i.e. without a universal scaling exponent as for conventional phase transitions. Based on a generic tight-binding (TB) model, we demonstrate this linear scaling relation to define the critical TPT transition point from the atomic limit to topological solid, regardless of lattice symmetry, periodicity and form of electron hopping. We validate this linear scaling by calculating TPTs in various 2D crystalline lattices (oblique, trigonal, square, rectangle, rhombic, etc.) as well as quasicrystalline lattices. Furthermore, we demonstrate this linear scaling sets an upper bound for the degree of disorder to destroy the topological order in a crystal by the case studies of vacancy formation and thermal disorder.

## 2. Model

Our TB model consists of three orbitals ($s$, $p_x$ and $p_y$) per site,

$$H = \sum_{i\alpha} \varepsilon_\alpha c^\dagger_{i\alpha} c_{i\alpha} + \sum_{\langle i\alpha, j\beta \rangle} t_{i\alpha, j\beta} c^\dagger_{i\alpha} c_{j\beta} + i\lambda \sum_i (c^\dagger_{ip_y} \sigma_z c_{ip_x} - c^\dagger_{ip_x} \sigma_z c_{ip_y}), \quad (1)$$

where $c_{i\alpha}^\dagger = (c_{i\alpha\uparrow}^\dagger, c_{i\alpha\downarrow}^\dagger)$ are electron creation operators on the $\alpha$ (= $s$, $p_x$, $p_y$) orbital at the $i$-th site and $\varepsilon_\alpha$ is the on-site energy of the $\alpha$ orbital. $t_{i\alpha,j\beta} = t_{\alpha\beta}(r_{ij})$ is the hopping integral. $\lambda$ is the SOC strength and $\sigma_z$ is the Pauli matrix. We set a cutoff distance $r_{cut}$ beyond which the hopping vanishes. Within the cutoff, $t_{\alpha\beta}(r_{ij}) = \text{SK}[\hat{r}_{ij}, V_{\alpha\beta\delta}(r_{ij})]$ follows the Slater-Koster scheme [23]. The radial dependence of the bond integral $V_{\alpha\beta\delta}(r_{ij})$ (with $\delta = \sigma \text{ or } \pi$) is captured by power-law or exponential decay functions in different materials [24, 25]. Since only the band inversion between $s$ and $p$ states of different parities is important for TPT, we focus on the 2/3 filling of electronic states hereafter (see Supplementary Material for details).

We define the average bond length $L$ for a system with $N$ atomic sites as,

$$L = \frac{1}{N} \sum_{\langle i,j \rangle} r_{ij}, \qquad (2)$$

where the summation runs over all the bonds within the cutoff (*i.e.*, $r_{ij} < r_{cut}$). It is worth noting that this expression is applicable to both crystalline and noncrystalline lattices, as discussed later.

### 3. Linear scaling of TPT

As shown in Figure 1, the critical point of a TPT can be roughly determined by a critical bandwidth $W_c = \frac{1}{2}(W_s + W_p - 2\lambda) = \Delta_{sp} - \lambda$, which depends only on the given atomic levels and SOC strength, independent of lattice types. According to the deformation-potential theory [26, 27], the energy levels under strain are expressed as

$$E_{band}(\epsilon_{ij}) = E_0 + \sum_{ij} \left(\frac{dE}{d\epsilon_{ij}}\right)\epsilon_{ij} + \cdots, \qquad (3)$$

where $\Xi_{v/c} = \frac{dE_{v/c}}{d\epsilon_{ij}}$ are the deformation potentials for electrons in the valance and conduction bands, respectively. Similarly, we linearize the band evolution process with the bandwidth evolving with the average bond length as

$$W_{s,p} = W_{s,p}^0 + \left(\frac{dW_{s,p}}{dL}\right)L + \cdots. \qquad (4)$$

Then, the energy gap is given by

$$E_g = \Delta_{sp} - \frac{1}{2}(W_s + W_p) - \lambda = E_g^\infty - \left(\frac{dE_g}{dL}\right)L + \cdots, \qquad (5)$$

where $\gamma = \frac{dE_g}{dL}$ effectively the electron hopping potential which is related to the band-edge deformation potentials. Within the TB approximation, the bandwidth $W$ of different lattices is approximatively proportional to the summation of nearest-neighbor hopping integrals [1], which changes with the average bond length $L$ in Eq. (2). Then, the critical transition point $L_c$ of TPT where the gap closes ($E_g = 0$) is simply determined by

$$L_c \approx \frac{E_g^\infty}{\gamma} + \cdots. \qquad (6)$$

Namely, the critical bond length $L_c$ is linearly proportional to the reciprocal of $\gamma$. As the average bond length $L$ is defined within the cutoff as presented in Eq. (2), the critical behavior of TPTs is mainly determined by the neighboring environment due to the "nearsightedness" of quantum-mechanical interactions [28].

4. **TPTs in crystal and quasicrystal lattices**

   To validate the above hypothesis, we first systemically calculated TPTs in various 2D periodic lattices. A trigonal lattice with an *sp* basis was found previously to host TI state, such as in Au/GaAs(111) and Bi/Si(111) systems [29, 30]. By tuning the average bond length $L$, the trigonal lattice undergoes a TPT between a NI and a TI state accompanied by an energy gap closing and reopening. Figure 2(a) shows the critical values of the TPT $L_c = 3.089$ Å for the trigonal lattice. In Figure 2(b), the orbital-resolved band structure indicates a nontrivial electronic topology beyond the TPT. It exhibits a band inversion around the Γ point between the *s*-orbital-derived conduction band and the *p*-orbital-derived valance band. The calculated $Z_2 = 1$, which is obtained by directly tracing the evolution of 1D hybrid Wannier charge center [31], confirming the TI state in this region. Furthermore, similar TPTs have been found in various typical 2D lattices including oblique (monoclinic), rectangular (orthorhombic), rhombic or centered rectangular (orthorhombic), trigonal (hexagonal), square (tetragonal), honeycomb, Lieb and semiregular Archimedean lattices (see Figure S3-S4 in Supplementary Material). For each lattice, we calculated the phase evolution diagram and determined $L_c$.

   Remarkably, we found that for more than 60 different lattices, $L_c$ exhibits a linear scaling with the reciprocal of $\gamma$, as shown in Figure 3. Numerical fitting gives a slope of $E_g^\infty = 1.5$ eV, which implies that Eq. (6) is valid independent of specific lattice symmetries. We also check the results using an exponentially decay function for electron hopping, which importantly confirms that the linear scaling is valid for different hopping potentials (i.e. materials) albeit with a different slope (see also Figure S8 in Supplementary Material). Thus, it is generally applicable but without a "universal" slope. We emphasize that the calculations cover almost all kinds of 2D crystalline lattices with different symmetries. More interestingly, the linear scaling is even applicable to quasicrystal lattices, as demonstrated by examples of Penrose-type pentagonal [20, 21] and Ammann-Beenker-type octagonal [22] quasicrystalline lattices, as also shown in Figure 3. This points to a general linear scaling of TPT in all the 2D systems, regardless of not only lattice symmetry but also periodicity.

5. **TPTs in crystals with disorder**

   As the definition of $\gamma$ and $L$ is the same for both crystalline and noncrystalline systems, one expects the linear scaling to be also applicable to define TPT in crystals with disorder. It is well known that the conducting edge state of a TI is distinguished from a normal metallic state by topological protection, so that the former is robust against non-magnetic "edge" disorder (defects or impurities). It is rooted in the bulk-boundary correspondence of a TI phase, as edge disorder cannot destroy bulk band topology. However, if bulk disorders occurred in a TI, bulk band topology and hence topological edge state could be destroyed. Thus, an intriguing and practically useful question is how robust a TI can be against bulk disorder? Below we will answer this question by applying the linear scaling of TPT to 2D crystals with two kinds of possible bulk disorder, formation of vacancies and thermal displacements.

We again considered a trigonal lattice with random vacancies in a wide range of concentration $\eta$ [see Figure 4(a) for example], and studied the TPT induced by decreasing $L$. As shown in Figure 4(b), TPTs between NI and TI states may occur for different vacancy concentration $\eta$, and the critical point $L_c$ decreases with increasing $\eta$. Correspondingly, the region of the TI phase becomes smaller with increasing $\eta$ and eventually disappears beyond a critically large $\eta_c$. Figure 4(c) shows the phase diagram in the $L$-$\eta$ parameter space. Apparently, the NI and TI phases are divided by a curve of zero energy gap. In order to confirm the TPT, we calculated the spin Bott index, a topological invariant we developed recently for TI systems with disorder [20–22]. It is found that there is a concomitant sharp jump in the spin Bott index $B_s$ across the phase boundary, confirming a TPT.

In addition, there is a large region of parameters in $\eta$ and $\gamma$ where the system is gapless, as shown in Figure 4(c). An important point is that there is a critical vacancy concentration $\eta_c$ below which a TI phase can only exist. The TI region shrinks and disappears at $\eta_c \approx 0.2$, which defines an upper bound for TPT in a 2D trigonal lattice with vacancies. This is expected to be a general phenomenon although the precise value of $\eta_c$ depends on specific model parameters. For $\eta < \eta_c$, the system is driven from a NI into a gapless phase through the intermediate TI region with the decreasing $L$; while for $\eta > \eta_c$, there is no TI region no matter how small $L$ is. We also investigated the samples with different sizes and found similar phase transitions. This confirms the applicability of the linear scaling of TPT in the thermodynamic limit of infinite lattice size.

We next investigate the effect of thermal disorder in destroying the topological phase in a 2D crystal. Due to thermal fluctuation, the interatomic distance $r_{ij}$ varies locally, which broadens the discrete peaks of the radial distribution function. It is noted that the melting transition from perfect crystalline to paracrystalline [32] and amorphous lattices [33] with increasing thermal fluctuation can be complicated. As an illustrative example, we adopted the quasi-lattice model [33] which assumes that the atomic displacements (**u**) away from their equilibrium positions follow a Gaussian distribution:

$$p(\boldsymbol{u}) = \frac{1}{\sqrt{2\pi\sigma^2}} exp(-\frac{u^2}{2\sigma^2}). \tag{7}$$

The mean-squared displacement $\sigma^2$, which represents the strength of thermal fluctuation, is approximately proportional to temperature $\sigma^2 \propto k_{\text{B}}T$ according to the compressibility equation [34, 35]. By increasing temperature $T$, the lattice transforms from a crystal to an amorphous gradually. We studied the TPT in trigonal lattices with thermal fluctuation-induced bond disorder [see Figure 4(d)]. As shown in Figure 4(e), the energy gaps $E_g$ for the TI region decrease and eventually disappear with increasing $\sigma$, indicating that the thermal disorder can actually destroy the nontrivial topology. Surprisingly, $L_c$ of TPT increases with increasing $\sigma$. Figure 4(f) shows the phase diagram of the thermal disorder system in the $L$-$\sigma$ parameter space. The NI and TI states are separated by a curve of zero energy gap. In strongly-disordered region, i.e., $\sigma > 0.15$ which represents an upper bound of thermal disorder for TPT, the intermediate TI phase disappears and the phase transition occurs between a NI and a gapless state directly. According to the Lindemann melting criterion the solid melts when $\sigma$ exceeds a threshold value (typically between 5% and 20% of the NN distance) [36]. Therefore, our results suggest the persistence of topological states even above melting temperature, i.e., the possible existence of "topological liquid".

Finally, we add the critical transition points $(1/\gamma, L_c)$ for TPTs of both random-vacancy and thermal-disorder lattices into Figure 3. Remarkably, they follow the same linear scaling relation as obtained above for various crystal and quasicrystal lattices.

6. **Discussion**

We note that the linear scaling as discovered by TB calculations and analyses from the "atomic limit" to TI is physically related to the well-known concept of deformation potential [26, 27] underlying the linear dependence of semiconductor band gap on external strain. This implies that by measuring the band gap as a function of strain (a manifestation of average bond length $L$) for a chosen semiconductor, one will be able to predict the critical point for TPT in the band evolution diagram (Figure 1) as well as how much strain is needed to convert this semiconductor into a TI. We are currently exploring this interesting possibility. Furthermore, the linear scaling will help us to better understand the physical nature of TPTs in terms of local atomic environment. There exists a critical atomic density below which the average bond length is too large so that the topological state would never occur [18, 19]. Experimentally, our finding suggests that topological states can be quite robust against a high degree of structural disorder that usually occurs during a non-equilibrium growth process. This may significantly ease the fabrication of topological materials for practical applications.

Finally, it is interesting to note that for crystals one can identify whether it is a TI by symmetry analysis of band topology to determine if TPT has occurred by linking the solid crystal to the atomic crystal assumed with the same symmetry, which have provided a powerful method to discover topological crystals [37–39]. However, our studies show that the TPT occurs regardless of symmetry, i.e., symmetry is a convenient means to determine the TPT in crystals, but not a mandatory condition. In addition, with certain crystalline symmetry constraint, topological semimetals can also occur around the critical point in the band evolution diagram we proposed. Although the illustrative calculation in this work is based on a specific model, the essential band inversion mechanism is not limited to the $s$ and $p$ orbitals, and other types of band inversion mechanisms between different orbitals are also feasible to achieve similar phase evolution diagrams. The unified view we provide here is applicable to any TPT that is triggered by a band-inversion mechanism, such as 2D/3D NI to topological (crystalline) insulator and 2D NI to Chern insulator transitions. However, it should not apply to TPTs that do not involve band inversion, such as quantum Hall effect hosted by a topological flat band [40] or a transition from a Dirac semimetal to TI upon opening a SOC gap like in graphene. Our discovery may also shed lights on understanding topological effects in other fields such as topological photonics [41], phononics [42], mechanics [43], metamaterials [44], and topolectrical circuits [45].

**Acknowledgments**

This work was supported by U.S. DOE-BES (Grant No. DE-FG02-04ER46148). This research used resources of the CHPC at the University of Utah and the National Energy Research Scientific Computing Center, a DOE Office of Science User Facility supported by the Office of Science of the U.S. Department of Energy under Contract No. DE-AC02-05CH11231.

**Author Contributions**
H. Huang and F. Liu designed the project. H. Huang performed theoretical calculation, and all authors prepared the manuscript.

**Conflicts of Interest**
The authors declare that there is no conflict of interest regarding the publication of this article.

**Data Availability**
All data needed to evaluate the conclusions in the paper are present in the paper and/or the Supplementary Materials. Additional data related to this paper may be requested from the authors.

**Supplementary Materials**
Supplementary Text: general band evolution; Details of methodology; TPTs of two-dimensional crystals and quasicrystals; random vacancy; thermal fluctuation; validity of the linear scaling for different form of electron hopping.
Figure S1: Schematic illustration of topological phase transition.
Figure S2: The calculation of formation of band structure from discrete levels of isolated atoms by decreasing average bong length.
Figure S3: Five 2D Bravais lattice.
Figure S4: The eight lattices based on semiregular Archimedean tilings.
Figure S5: Several decorated trigonal lattices.
Figure S6: Atomic model of the Penrose-type and the Ammann-Beenker-type quasicrystal lattices.
Figure S7: The linear scaling relation in various 2D crystalline lattices using the power-law decay function ($1/r_{ij}^2$).
Figure S8: The linear scaling relation in various 2D crystalline lattices using the exponentially decay function ($e^{-2.3(r_{ij}-1)}$).
Table S1. The electron hopping potential $\gamma$ and the critical bond length $L_c$ for TPT in typical 2D crystals and quasicrystal lattices.

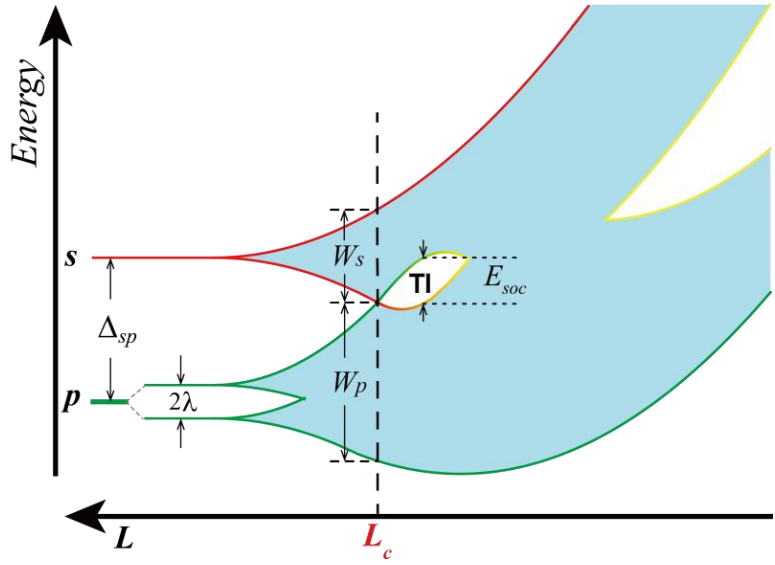

FIGURE 1: *Schematic illustration of TPT in band evolution diagram.* By decreasing average bond length $L$, the bandwidth increases gradually and a TPT occurs at $L_c$, which is inversely proportional to the electron hopping potential $\gamma = dE_g/dL$.

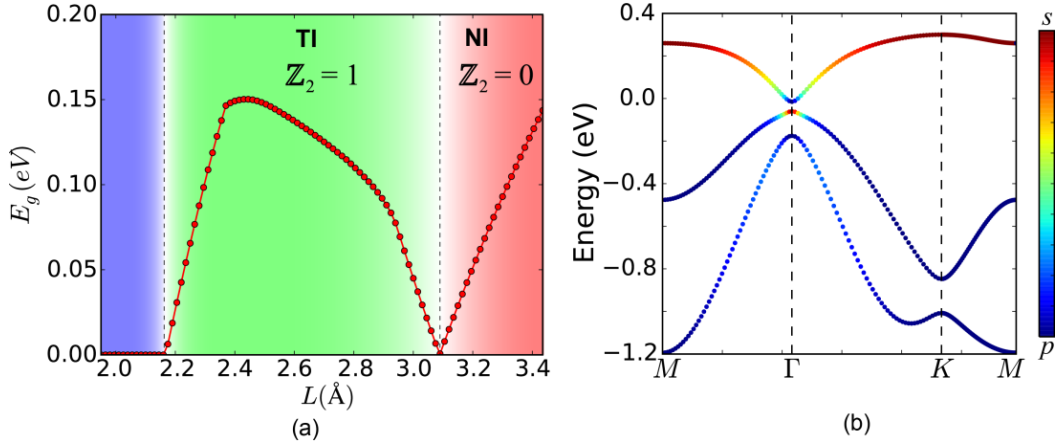

FIGURE 2: *TPT in a trigonal lattice.* The parameters used here are $\varepsilon_s = 0.18, \varepsilon_p = -0.65, \lambda = 0.08, V_{ss\sigma} = -0.04, V_{sp\sigma} = 0.09, V_{pp\sigma} = 0.18$ and $V_{pp\pi} = 0.005$ eV at lattice constant $a = 1$Å. The color of dots represents the relative weight of $s$ and $p$ orbitals. (a) Energy gap $E_g$ and $Z_2$ index versus average bond length $L$. A TPT between a NI and a TI is clearly visible. (b) Band structure of the trigonal lattice at the TI phase.

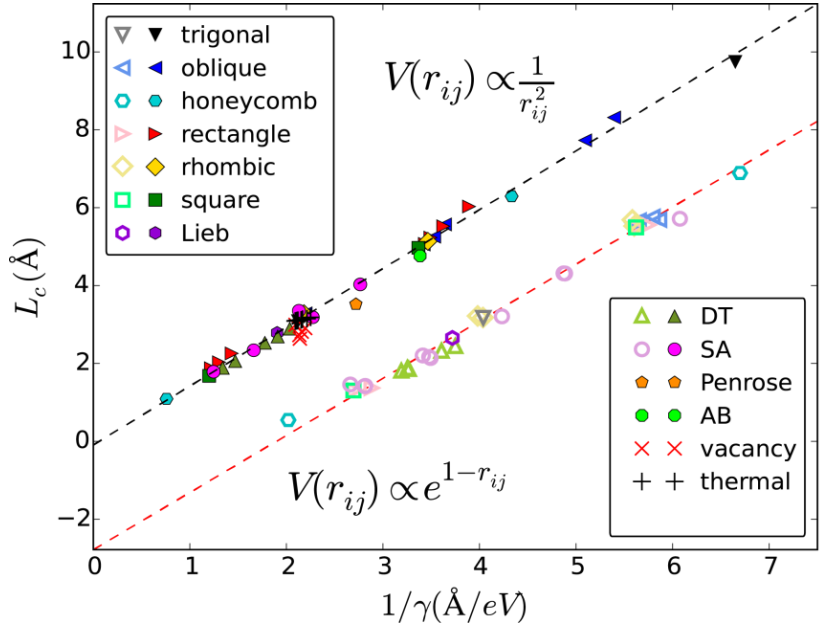

FIGURE 3: *Linear scaling of TPT.* The linear scaling relation between the critical value of average bond length $L_c$ and the reciprocal electron hopping potential ($1/\gamma$) for TPT in all the studied 2D periodic, quasicrystalline and disorder lattices, including oblique (monoclinic), rectangular (orthorhombic), rhombic or centered rectangular (orthorhombic), trigonal (hexagonal), square(tetragonal), honeycomb, Lieb, decorated-trigonal (DT), semiregular-Archimedean (SA), Penrose-type and Ammann-Beenker-type (AB) lattices (see Figure S4 in Supplementary Material). The red "_" and black "+" denote trigonal lattices with random vacancies and thermal disorder, respectively. The data marked by filled (open) symbols are calculated using the power-law (exponentially) decay function for the radial dependence of electron hoppings.

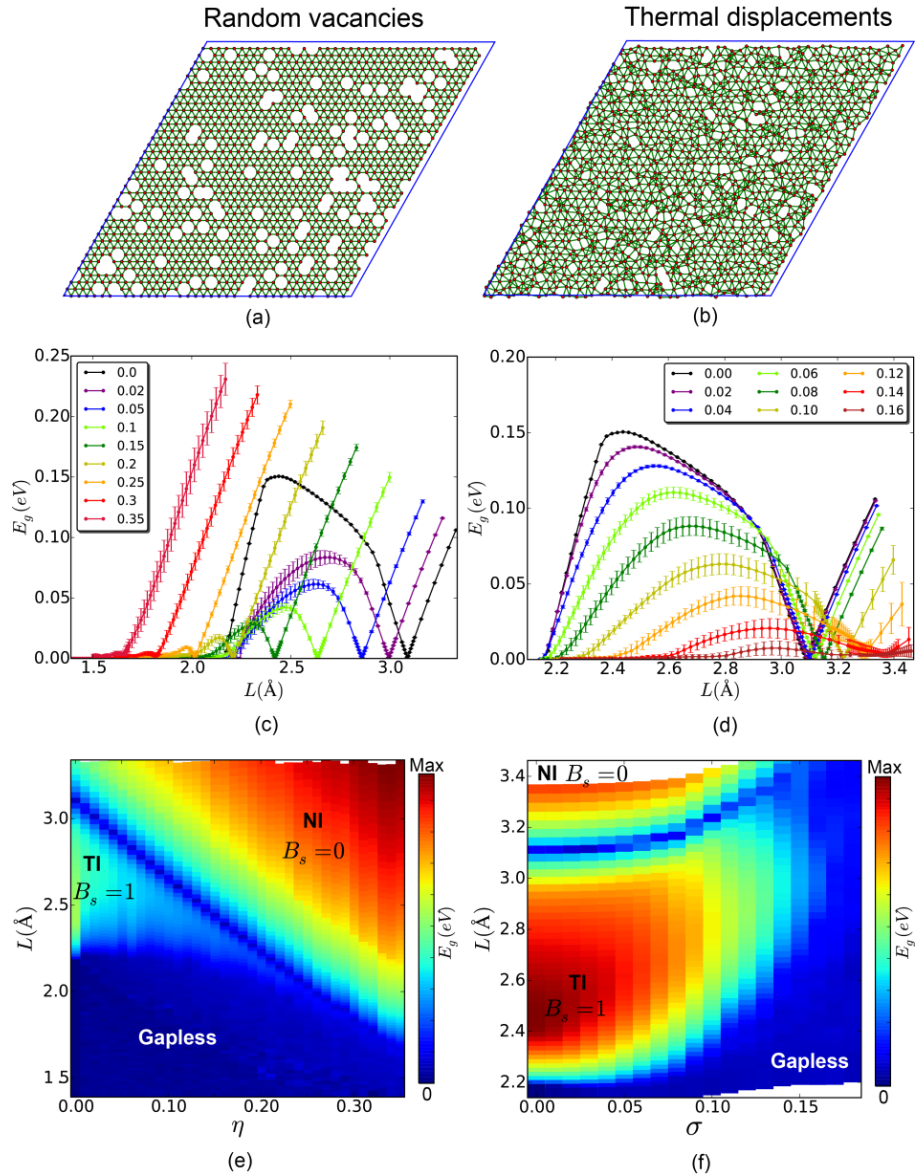

FIGURE 4: *TPT in crystals with disorders*. Atomic configuration of a trigonal lattice (a) with random vacancies at $\eta = 0.15$, and (b) with thermal disorder at $\sigma = 0.16$. Energy gap $E_g$ versus $L$ for samples (c) with difference $\eta$, and (d) with different $\sigma$. Phase diagram of trigonal lattices (e) with random vacancies in the parameter space of $\eta$ and $L$, and (f) with thermal disorder in the parameter space of $\sigma$ and $L$. The color represents the size of energy gap.